\documentstyle{elsart}
\input psfig

\begin{document}
\begin{frontmatter}

\title{How Uncertain Are Solar Neutrino Predictions?}
\author{\underline{John N. Bahcall}\thanksref{jnbemail}} 
and \author{Sarbani Basu\thanksref{sbemail}}
\address{School of Natural Sciences, 
Institute for Advanced Study, Princeton, New Jersey 08540}
\author{M. H. Pinsonneault\thanksref{mhpemail}}
\address{Department of Astronomy, Ohio State University, Columbus, Ohio 43210}
\thanks[jnbemail]{E-mail address: jnb@sns.ias.edu}
\thanks[sbemail]{E-mail address: basu@sns.ias.edu}
\thanks[mhpemail]{E-mail address: pinsono@payne.mps.ohio.state.edu}

\begin{abstract}
Solar neutrino fluxes and 
sound speeds are calculated using 
a  systematic reevaluation of nuclear fusion rates. 
The largest uncertainties are identified  
and their effects on the solar neutrino fluxes are estimated. 
\end{abstract}

\end{frontmatter}

Five solar neutrino experiments (chlorine, Kamiokande, GALLEX, SAGE,
and Super-Kamiokande) have measured solar neutrinos with
approximately the fluxes and energies predicted by standard solar
models, confirming empirically the basic picture of stellar energy
generation. However,
robust quantitative  differences exist between the neutrino
experiments and   the combined 
predictions of minimal standard electroweak theory and stellar evolution
models.  Many authors have suggested that these results provide the
first evidence of physics beyond the minimal standard electroweak
model.  

What are the principal uncertainties in the standard model
predictions?   In this paper, we determine
the uncertainties in the solar neutrino calculations that arise from errors
in the nuclear fusion cross sections and show that these uncertainties, while
relatively small,  are 
currently the 
largest sources of recognized errors in the neutrino predictions.

In January, 1997, the Institute for Nuclear Theory (INT) hosted a workshop
devoted to determining the best estimates and
the uncertainties in the
most important solar fusion reactions. Thirty-nine experts in 
low energy nuclear
experiments and theory, representing many different
research groups and points of view,  participated in the workshop and
evaluated the existing experimental data and theoretical 
calculations. Their conclusions  have been
summarized in a detailed article authored jointly by the participants
and to be published by 
the Reviews of Modern Physics \cite{adelberger98}.  In general 
outline, the conclusions
of the INT workshop paper confirmed and strengthened 
previous standard analyses of
nuclear fusion rates, although in a few important  cases 
(for the ${\rm ^3He}(\alpha, \gamma){\rm ^7Be}$, 
${\rm ^7Be}(p, \gamma){\rm ^8B}$, and 
${\rm ^{14}N}(p, \gamma){\rm ^{15}O}$ reactions)
the estimated
uncertainties were determined to be larger than previously believed.

\begin{table}[t]
\baselineskip=16pt
\centering
\begin{minipage}{3.5in}
\caption[]{Standard Model Predictions (BP98): 
solar neutrino fluxes and neutrino capture rates, with $1\sigma$
uncertainties from all sources (combined quadaratically).\protect\label{bestestimate}}
\begin{tabular}{llcc}
\noalign{\smallskip}
\hline
\noalign{\smallskip}
Source&\multicolumn{1}{c}{Flux}&Cl&Ga\\
&\multicolumn{1}{c}{$\left(10^{10}\ {\rm cm^{-2}s^{-1}}\right)$}&(SNU)&(SNU)\\
\noalign{\smallskip}
\hline
\noalign{\smallskip}
pp&$5.94 \left(1.00^{+0.01}_{-0.01}\right)$&0.0&69.6\\
pep&$1.39 \times 10^{-2}\left(1.00^{+0.01}_{-0.01}\right)$&0.2&2.8\\
hep&$2.10 \times 10^{-7}$&0.0&0.0\\
${\rm ^7Be}$&$4.80 \times 10^{-1}\left(1.00^{+0.09}_{-0.09}\right)$&1.15&34.4\\
${\rm ^8B}$&$5.15 \times 10^{-4}\left(1.00^{+0.19}_{-0.14}\right)$&5.9&12.4\\
${\rm ^{13}N}$&$6.05 \times
10^{-2}\left(1.00^{+0.19}_{-0.13}\right)$&0.1&3.7\\
${\rm ^{15}O}$&$5.32 \times
10^{-2}\left(1.00^{+0.22}_{-0.15}\right)$&0.4&6.0\\
${\rm ^{17}F}$&$6.33 \times
10^{-4}\left(1.00^{+0.12}_{-0.11}\right)$&0.0&0.1\\
\noalign{\medskip}
&&\hrulefill&\hrulefill\\
Total&&$7.7^{+1.2}_{-1.0}$&$129^{+8}_{-6}$\\
\noalign{\smallskip}
\hline
\noalign{\smallskip}
\end{tabular}
\end{minipage}
\end{table}

The purpose of this article is to present calculations of solar
neutrino fluxes 
and solar sound velocities, with special attention to their uncertainties,
that were made using 
the recommended INT nuclear reaction rates and the best available
other input data.

Our results can be compared directly with the 
observed rates in 
solar neutrino experiments and be used as input 
for detailed analyses of the particle
physics implications of the measured solar neutrino rates.
We identify the most 
important nuclear parameters that need to be measured more accurately 
in laboratory experiments and
determine the precision that is required.
By comparing our solar models with five recent, precise 
helioseismological determinations of the sound velocities, we estimate
the size of the 
remaining errors in the model
calculations.

Table~1 gives the  neutrino fluxes and their 
uncertainties for our best standard solar model (hereafter BP98).
The solar model
makes use of the INT nuclear reaction rates \cite{adelberger98}, 
recent (1996) Livermore
OPAL opacities \cite{opacity}, 
the OPAL equation of state \cite{eos}, and 
electron and ion screening as indicated by recent
calculations \cite{gruzinov98}. 
The adopted uncertainties in input parameters are given in Table~2
and the associated text.
We have also made small improvements in our energy generation code,
which will be described in detail in a future publication.

The theoretical predictions in Table~1 
disagree with the observed neutrino event rates,
which are \cite{experiments}: 
$2.55 \pm 0.25$ SNU (chlorine), $73.4 \pm 5.7$ SNU (GALLEX and SAGE
gallium experiments), and $(2.80 \pm 0.19({\rm stat}) \pm
0.33({\rm syst}) ) \times 10^6 
{\rm cm^{-2} s^{-1} }$ ($^8$B flux from Kamiokande).

The principal differences between the results shown in Table~1 and the
results presented in our last systematic publication of calculated
solar neutrino fluxes \cite{BP95} is a  $1.3\sigma$ decrease in the 
$^8$B neutrino flux and $1.1 \sigma$ decreases
in the $^{37}$Cl and $^{71}$Ga capture rates.  These
decreases are due principally 
to the lower ${\rm ^7Be}(p,\gamma){\rm ^8B}$  cross section adopted
by Adelberger et al. \cite{adelberger98}.
If we use, as in our recent previous publications,  
the Caltech (CIT) value for the $^8$B production cross
section \cite{CIT}, then the $^8$B flux is 
$\phi\left({\rm ^8B, CIT}\right) = 6.1^{+1.1}_{-0.9} 
\times 10^6~{\rm cm^{-2}s^{-1}}$,
$
\Sigma\left(\phi\sigma\right)_i \big\vert\lower10pt
\hbox{\scriptsize Cl,~CIT} =
8.8^{+1.4}_{-1.1}\ \ {\rm SNU} $,  and 
$\Sigma\left(\phi\sigma\right)_i 
\big\vert\lower10pt\hbox{\scriptsize\rm Gallium} = 131^{+9}_{-7}\ \
{\rm SNU}$, all 
of which are within ten percent of the
Bahcall-Pinsonneault 1995 best-estimates.
The difference between the INT and the CIT estimates of the $^8$B
production cross section is due almost entirely to the decision by the
INT group to base their estimate on only one (the best documented) of
the six experiments analyzed by the CIT collaboration.

Table~2 summarizes the uncertainties in the most important
solar neutrino fluxes and in the Cl and Ga event rates 
due to different nuclear fusion reactions (the
first four entries), the heavy element to hydrogen mass ratio (Z/X),
the radiative opacity, the solar luminosity, the assumed solar age,
and the helium and heavy element diffusion coefficients. 
The ${\rm ^{14}N} + p$ reaction causes a
0.2\% uncertainty in the predicted pp flux and a 0.1 SNU uncertainty
in the Cl (Ga) event rates.

The predicted event rates for
the chlorine and gallium experiments use recent improved calculations
of neutrino
absorption cross sections \cite{nuabs}. The uncertainty in the
prediction for the gallium rate is dominated by  uncertainties in the
neutrino absorption cross sections, $+6.7$ SNU ($7$\% of the predicted
rate) and $-3.8$ SNU ($3$\% of the predicted rate). The
uncertainties in the chlorine absorption cross sections cause an
error, $\pm 0.2$ SNU ($3$\% of the predicted rate), that is relatively
small compared to other
uncertainties in predicting the rate 
for this experiment.  For non-standard neutrino energy
spectra that result from new neutrino physics, the uncertainties in
the predictions 
for currently favored solutions (which reduce the contributions from
the least well-determined $^8$B neutrinos) will
in general be less than the values quoted here for standard spectra
and must be calculated using the appropriate cross section uncertainty
for each neutrino energy \cite{nuabs}.

\begin{table}
\baselineskip=16pt
\centering
\begin{minipage}{5.5in}
\caption[]{Average uncertainties in neutrino fluxes and event rates 
due to different input data.  The flux uncertainties are expressed in
fractions of the total flux and the event rate uncertainties are
expressed in SNU.  The ${\rm ^7Be}$ electron capture rate causes an
uncertainty of $\pm 2\%$ \cite{be7paper} that affects only the ${\rm
^7Be}$ neutrino flux.  The average fractional uncertainties for
individual parameters are shown.  See text for discussion of
asymmetric uncertainties and uncertainties due to radiative opacity or
diffusion.}
\begin{tabular}{@{\extracolsep{-5pt}}lccccccccc}
\noalign{\smallskip}
\hline
\noalign{\smallskip}
$<$Fractional&pp&${\rm ^3He ^3He}$&${\rm ^3He ^4He}$&${\rm ^7Be} +
p$&$Z/X$&opac&lum&age&diffuse\\
uncertainty$>$&0.017&0.060&0.094&0.106&0.033&&0.004&0.004\\
\noalign{\smallskip}
\hline
\noalign{\smallskip}
Flux\\ \cline{1-1}
\noalign{\smallskip}
pp&0.002&0.002&0.005&0.000&0.002&0.003&0.003&0.0&0.003\\
${\rm ^7Be}$&0.0155&0.023&0.080&0.000&0.019&0.028&0.014&0.003&0.018\\
${\rm ^8B}$&0.040&0.021&0.075&0.105&0.042&0.052&0.028&0.006&0.040\\
\noalign{\medskip}
SNUs\\ \cline{1-1}
\noalign{\smallskip}
Cl&0.3&0.2&0.5&0.6&0.3&0.4&0.2&0.04&0.3\\
Ga&1.3&0.9&3.3&1.3&1.6&1.8&1.3&0.20&1.5\\
\noalign{\smallskip}
\hline
\noalign{\smallskip}
\end{tabular}
\end{minipage}
\end{table}

The nuclear fusion
uncertainties in Table~2 were taken  from Adelberger et
al. \cite{adelberger98}, 
the neutrino cross section uncertainties from \cite{nuabs}, 
the heavy element uncertainty was taken from
helioseismological measurements \cite{basu97}, the luminosity
and age uncertainties were adopted from BP95 \cite{BP95},  the
$1\sigma$ fractional uncertainty in the diffusion rate was taken to be
$15$\% \cite{thoul}, which is supported by helioseismological evidence
 \cite{prl97}, and the opacity uncertainty was determined by comparing
the results of fluxes computed using the older Los Alamos opacities
with fluxes computed using the modern Livermore opacities \cite{BP92}. 
 To include the effects of asymmetric errors, the code exportrates .f
(see below)
was run with different input uncertainties and the results averaged.

Many authors
have used the results of solar neutrino experiments and the calculated
best-estimates and uncertainties in the standard solar model fluxes to
determine the allowed ranges of neutrino parameters in  different
particle physics models. To systematize the calculations in the solar
neutrino predictions,
we have constructed an exportable 
computer code, exportrates.f, which evaluates the uncertainties in the
predicted neutrino fluxes and capture rates from all
the recognized sources of errors in the input data.  This code is
available at www.sns.ias.edu/$\sim$jnb (see Solar Neutrino Software
and Data); it contains a description of how each of the uncertainties
listed in Table~2 was determined and used. 

The low energy cross section
of the ${\rm ^7Be} + p$ reaction is the most important quantity that must be
determined more accurately in order to decrease the error in the
predicted event rates in solar neutrino experiments.
The $^8$B neutrino flux that is measured by the
Kamiokande \cite{experiments}, 
Super-Kamiokande \cite{totsuka}, and SNO \cite{sno} 
experiments is, in all standard solar model calculations, 
directly proportional to the ${\rm ^7Be} + p$ cross section. 
If the $1\sigma$ uncertainty in this cross section can be reduced
by a factor of two to 5\%, then it will no longer be the limiting
uncertainty in predicting the crucial $^8$B neutrino flux (cf. Table~2).

The ${\rm ^7Be}$ neutrino flux will be measured by BOREXINO \cite{borexino}.  
Table~2 shows
that the theoretical uncertainty in this flux is dominated by the
uncertainty in the measured laboratory rate for the ${\rm ^3He-^4He}$
reaction. In order that the uncertainty from the ${\rm ^3He-^4He}$ cross
section be reduced to
a level, 3\%,that is comparable to the uncertainties 
in calculating the $^7$Be flux that arise from
other sources, the fractional uncertainty in the ${\rm ^3He-^4He}$ 
low energy cross
section factor must be reduced to 3.5\%.  This goal is not easy, but
it appears to be 
 achievable.  The six published determinations of the low energy
cross section factor using measurements of the capture $\gamma$-rays currently
have a $1\sigma$ uncertainty of $3.2$\%, but they differ by
$2.5\sigma$ (14\%)
from the value determined by counting the $^7$Be
activity \cite{adelberger98}. 

Could the solar model calculations be wrong by enough to explain the
discrepancies between predictions and measurements for solar neutrino
experiments?  Helioseismology, which confirms predictions of the
standard solar model to high precision, suggests that the answer is
probably ``No.''

Fig.~1 shows the fractional differences between the most accurate 
available sound speeds
measured by helioseismology \cite{heliobest} and 
sound speeds calculated with our best
solar model (with no free parameters). 
The horizontal line corresponds to the hypothetical case in which the
model predictions exactly match the observed values.
The rms fractional 
difference between the
calculated and the measured sound speeds is $1.1\times10^{-3}$ for the
entire region over which the sound speeds are measured, $0.05R_\odot <
R < 0.95 R_\odot$.  In the solar core, $0.05R_\odot <
R < 0.25 R_\odot$ (in  which about $95$\% of the solar energy 
and neutrino flux is
produced in a standard model), the rms fractional 
difference between measured and
calculated sound speeds is $0.7\times10^{-3}$.

\begin{figure}[t]
\centerline{\psfig{figure=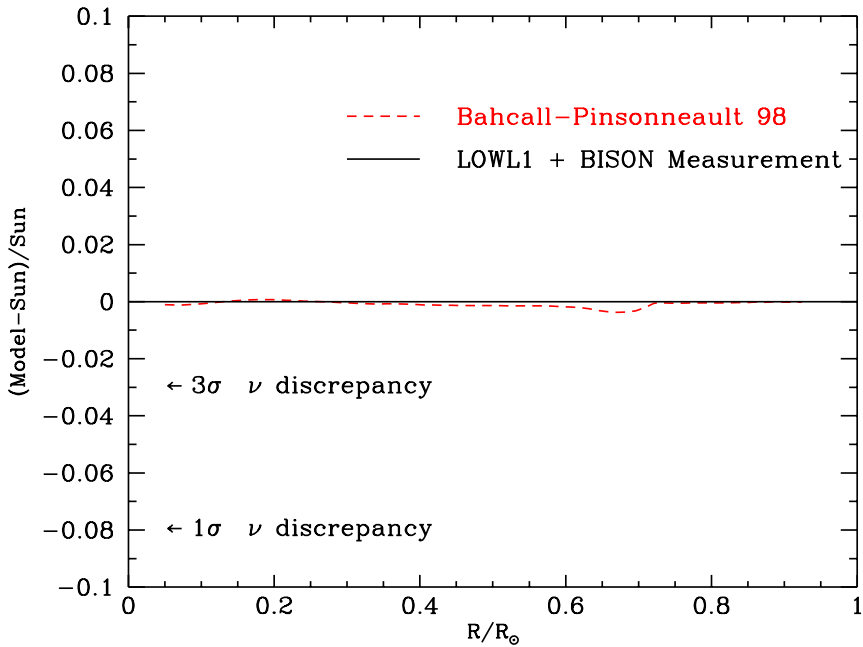,width=5in,angle=270}}
\caption[]{Predicted versus Measured Sound Speeds. 
This figure shows
the excellent agreement between the calculated (solar model BP98, Model) 
 and the measured (Sun) sound
speeds, a fractional difference of $0.001$ rms for all 
speeds measured between $0.05 R_\odot$
and $0.95 R_\odot$. The vertical scale is chosen so as 
to emphasize that the fractional error is much smaller than generic
changes in the model, $0.03$ to $0.08$, that might
significantly affect the solar neutrino predictions.
\protect\label{modelsunnu}}
\end{figure}

Helioseismological measurements also determine 
two other parameters that help characterize
 the outer part of the sun (far from
the inner region in which neutrinos are produced): the depth of the
solar convective zone (CZ), the region in the outer part of the sun
that is fully convective, and the present-day surface abundance by mass of
helium ($Y_{\rm surf}$).  The measured values, $R_{\rm CZ} =
(0.713 \pm 0.001) R_\odot$ \cite{basu95}, and $Y_{\rm surf} = 0.249 
\pm 0.003$ \cite{basu97},
are in satisfactory agreement with the values predicted by the solar
model BP98, namely, $R_{\rm CZ} =
0.714 R_\odot$, and $Y_{\rm surf} = 0.243$.
However, we shall see below that precision measurements of the sound
speed near the transition between the radiative interior (in which energy
is transported by radiation)  and the
outer convective zone (in which energy is transported by convection) 
reveal small discrepancies between the model
predictions and the observations in this region.

If solar physics were responsible for the solar neutrino problems, how
large would one expect the discrepancies to be between solar 
model predictions and
helioseismological observations?
The characteristic size of the discrepancies 
 can be estimated  using the results of the neutrino experiments and  
scaling laws for
neutrino fluxes and sound speeds.

All recently published solar models predict essentially 
the same fluxes from the fundamental
pp and pep reactions (amounting to $72.4$ SNU in gallium
experiments, cf. Table~1), which are closely related to the solar
luminosity.  
Comparing the measured and the standard predicted rate for the gallium
experiments, the $^7$Be flux must be reduced by a factor $N$
if the disagreement is not to exceed $n$ standard deviations, where 
$N$ and $n$ satisfy  $72.4 + (34.4)/N = 73.4 + n \sigma$. For a $1
\sigma $ ($3\sigma$) disagreement, $ N = 5.1 (1.9)$.  
Sound
speeds scale like the square root of the local temperature divided by
the mean molecular weight
and the $^7$Be neutrino flux scales approximately as the $10$th power
of the temperature \cite{ulmer}. 
Assuming that the temperature changes are dominant,
agreement to within $1\sigma$ would
require fractional changes of order $0.08$ in sound speeds ($3\sigma$
could be reached with $0.03$ changes), if all model changes were in the
temperature.
This argument is conservative because it ignores the contributions
from the $^8$B and CNO neutrinos which contribute to the observed
counting rate (cf. Table~1) and which, if included, would require an
even larger reduction of the $^7$Be flux. 

We have chosen the vertical scale in Fig.~1 to be appropriate for
fractional differences between measured and predicted 
sound speeds that are of order $0.03$ to $0.08$ and 
that might therefore affect solar neutrino calculations. 
Fig.~1 shows that the characteristic agreement between solar model
predictions and helioseismological measurements is more than a factor
of $30$ better than would be expected if there were a solar model
explanation of the solar neutrino problems.

Given the helioseismological measurements, how uncertain are the solar
neutrino predictions? We provide a tenative answer to this question by
examining more closely the small discrepancies shown in Fig.~1
between the observed and calculated sound speeds.

There has been an explosion of precise helioseismological data in the
last year.  Fig.~2 compares
the results of five different observational determinations of the sound
speeds in the sun with the results of our best solar model; the
helioseismological discussion in ref. \cite{prl97} made use of only the
LOWL1 data. The small features discrepancies shown in Fig.~2 are
robust; they occur when comparisons are made with all the data sets.
The vertical scale for Fig.~2 has been expanded by a factor of $20$ 
with respect to the scale of Fig.~1 in
order to show the small but robust discrepancies. References to the
different helioseismological measurements are given in the caption to
Fig.~2. 

\begin{figure}[t]
\centerline{\psfig{figure=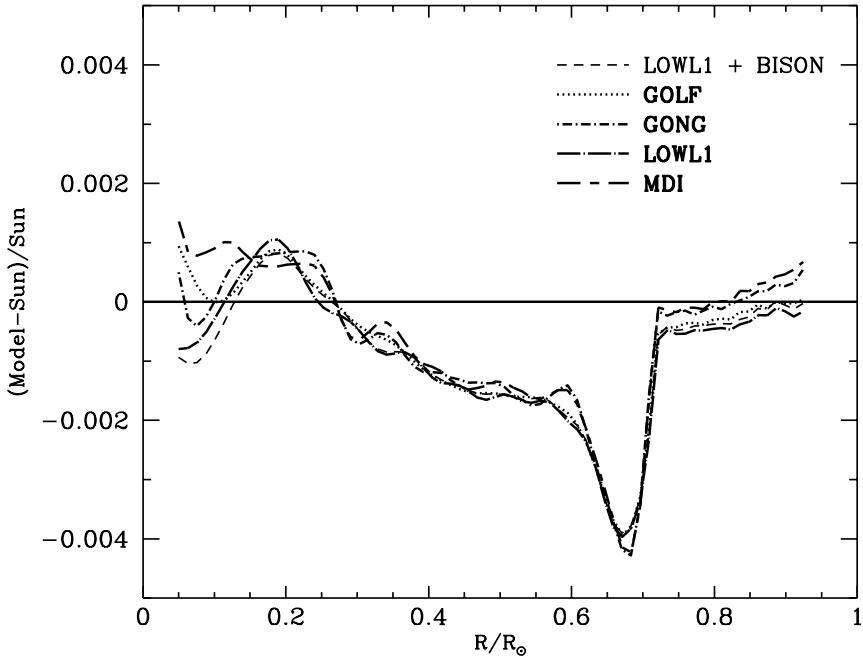,width=5in,angle=270}}
\caption[]{Five  Precise Helioseismological Measurements. The predicted
BP98 sound speeds are compared with five different helioseismological
measurements \cite{fivehelio}. 
The vertical scale has been expanded by a factor of $20$
relative to Fig.~1.\protect\label{modelsunhelio}}
\end{figure}

All five of the precise  helioseismological measurements show
essentially the same difference between the model and the solar sound speeds
near the  base of the convective zone.  The sharp edge to this
feature occurs near the present base of the convective zone at 
$R_{CZ} = 0.713 R_\odot$. 

What could be the cause of the broad feature shown in Fig.~2 that
stretches from about $0.3R_\odot$ to about $0.7R_\odot$? 
This feature may  be due to some combination of  small errors 
in the adopted radiative opacities or equation of state 
and the oversimplification we use
in our stellar evolution code of a sharp boundary between the
radiative and the convective zones. We can  make an 
estimate of the likely effect  of
hypothetical improved physics 
on the calculated neutrino fluxes
by considering what happens if we change
the adopted opacity.

\begin{figure}[t]
\centerline{\psfig{figure=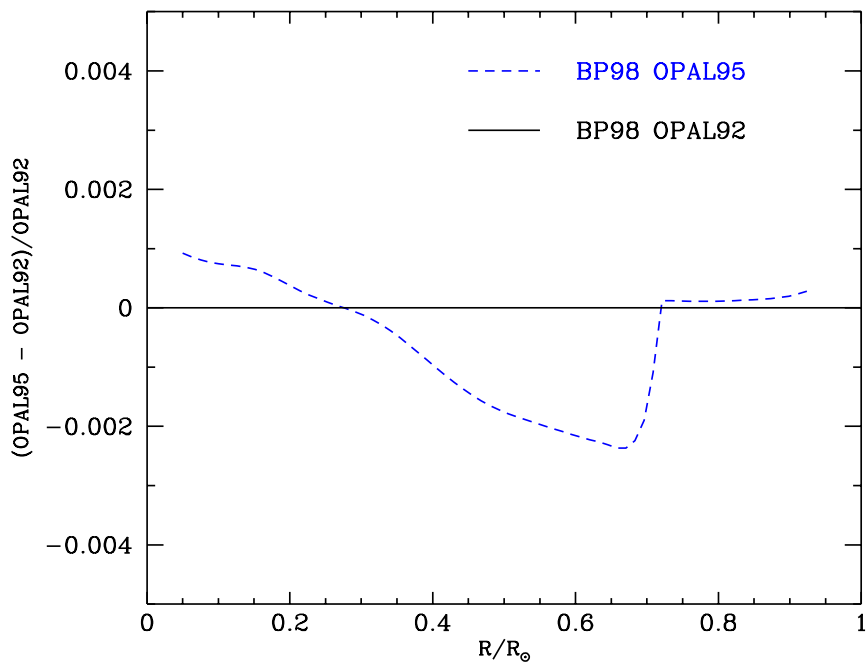,width=5in,angle=270}}
\caption{The effect of opacity on the calculated sound speeds.  The
figure shows the difference between the calculated sound speeds for
two solar models that differ only in the version used of the OPAL
opacities \cite{opacity}, 1992 or 1995.}
\end{figure}

Fig.~3 shows the fractional 
difference in the computed sound speeds obtained from
two solar models that are identical except for the adopted radiative
opacity \cite{opacity},
either the version from OPAL92 or the later version from OPAL95. It is
apparent from Fig.~3 
that the difference in sound speeds caused by using the OPAL95
opacity rather than OPAL92 opacity produces a feature that is similar
in shape and in magnitude to the broad feature in Fig.~2.  
For $0.3 < R/R_\odot < 0.6$, the OPAL95 opacity is about $2$\% less
than the OPAL92 opacity for the conditions in the BP98 model. The
opacity difference increases to about $6$\% near the base of the
convective zone. 
A change in opacity of about the same size but opposite
in sign to that which occurred between OPAL92 and OPAL95 
would remove most of the broad discrepancy, but at the price of
producing a slightly deeper convective zone, $0.7105 R_\odot$, than is
observed.

What are the implications for solar neutrino predictions of the
existence of the broad $0.2$\% discrepancy highlighted in Fig.~2?
Table~3 shows the neutrino fluxes predicted by the BP98 model 
constructed using OPAL92
opacities. The chlorine rate is increased relative to the standard
BP98 model (cf. Table~1) by $5$\% and the $^7$Be and $^8$B fluxes are
increased by $3$\% and $5.5$\% , respectively.  These changes are 
plausible estimates of the changes in the neutrino rates that
may be anticipated from further precision improvements in solar
models in the outer radiative zone between $0.3 R_\odot$ and $0.7
R_\odot$. 

The narrow feature near the base of the convective zone may be caused by
mixing that is related to the observed \cite{marc} depletion of Li and Be.
Several independent calculations \cite{several} of the required amount of
mixing indicate that the effect on the neutrino fluxes is small, less
than a 1\% change in the pp flux, and a $2$\% ($4$\%) decrease in the 
$^7$Be ($^8$B) neutrino flux.

There is a  smaller difference between the models and the
observations centered near $R = 0.2 R_\odot$.  
This feature occurs using all available helioseismological data and is
robust against changes in the method of inversion and inversion
parameters.
This feature could result, for example, from 
a $2$\% inaccuracy in the Livermore opacity at $9\times10^6$
K \cite{tripathy}
or a $0.1$\% inaccuracy in the OPAL 
evaluation of the adiabatic index $\Gamma_1$.
Detailed calculations \cite{bahcall69} 
of the sensitivity of the neutrino fluxes to
changes in opacity or equation of state suggest that either of the changes
mentioned above would affect the most sensitive calculated 
neutrino fluxes by about $2$\%.

\begin{table}[t]
\centering
\begin{minipage}{2.5in}
\caption{Standard solar model with OPAL92 opacities: solar neutrino
fluxes. The predicted chlorine capture rate is $8.1$ SNU and the
gallium rate is $131$ SNU.}
\begin{tabular*}{2.5in}[]{@{\extracolsep{\fill}}lc}
\noalign{\smallskip}
\hline
\noalign{\smallskip}
Source&Flux\\
&($10^{10}~{\rm cm^{-2}s^{-1}}$)\\
\noalign{\smallskip}
\hline
\noalign{\smallskip}
pp&5.92\\
pep&$1.39 \times 10^{-2}$\\
hep&$2.08 \times 10^{-7}$\\
${\rm ^7Be}$&$4.94 \times 10^{-1}$\\
${\rm ^8B}$&$5.44 \times 10^{-4}$\\
${\rm ^{13}N}$&$6.25 \times 10^{-2}$\\
${\rm ^{15}O}$&$5.52 \times 10^{-2}$\\
${\rm ^{17}F}$&$6.59 \times 10^{-4}$\\
\noalign{\smallskip}
\hline
\noalign{\smallskip}
\end{tabular*}
\end{minipage}
\end{table}

In conclusion, we note that three decades of refining the input data
and the solar model calculations has led to a predicted standard model
event rate for
the chlorine experiment, $7.7$ SNU, which is very close to the
best-estimate value 
obtained in 1968 \cite{bahcall68}, which was $7.5$ SNU. The
situation regarding solar neutrinos is, however, completely different
now, thirty years later.
 Four experiments have confirmed the detection of solar neutrinos.
Helioseismological measurements show (cf. Fig.~1) that hypothetical
deviations from the standard solar model that seem to be  required
by simple scaling laws  
to fit just the 
gallium solar neutrino results are at least a factor of $30$ larger
than the rms disagreement between the standard solar model predictions and the
helioseismological 
observations. This conclusion does not make use of 
the additional evidence which
points in the same direction from the chlorine, Kamiokande, and
SuperKamiokande experiments.
The comparison between observed and calculated 
helioseismological sound speeds is now
so precise ($\sim 0.1$\% rms) that  Fig.~2 indicates the need for
an improved physical description of the broad region
between $0.3 R_\odot$ and $0.7 R_\odot$.  The indicated improvement
may increase the $^7$Be and $^8$B neutrino fluxes by $\sim 5$\%
(cf. Table~1 and Table~3). The  narrow deep feature near $0.7 R_\odot$
suggests that mixing, possibly associated with Li depletion, 
 might reduce the $^7$Be and $^8$B 
neutrino fluxes by somewhat less than $5$\% \cite{several}.
Measurements of the low energy cross sections for the 
${\rm ^3He}(\alpha, \gamma){\rm ^7Be}$ reaction to a $1\sigma$ accuracy
of $3$\% and of the ${\rm ^7Be}(p, \gamma){\rm ^8B}$ reaction to an accuracy of
$5$\% are required in order that uncertainties in these laboratory
experiments  not limit the
information that can be obtained from 
solar neutrino experiments  about the solar interior  and 
about  fundamental neutrino physics.

JNB acknowledges support from NSF grant \#PHY95-13835.

\end{document}